\documentclass[useAMS,usenatbib]{mn2e}
\usepackage{graphicx}
\usepackage{times}

\def\kms{km ${\rm s}^{-1}$}

\def\ch2{$\chi^2$}

\def\kms {\hbox{${\rm km\ s}^{-1}$}}

\def\scm  {$\hbox{{\rm cm}}^{-2}$}    %cm-2
\def \AL {$\alpha $}     %  gr. alpha
\def \HI {H{\sc \,i}}

\def\lapp{\ifmmode\stackrel{<}{_{\sim}}\else$\stackrel{<}{_{\sim}}$\fi}
\def\gapp{\ifmmode\stackrel{>}{_{\sim}}\else$\stackrel{>}{_{\sim}}$\fi}

\title[21-cm in Mg{\sc \,ii} absorbers]{On the detectability of \HI\ 21-cm in Mg{\sc \,ii} absorption systems}
  \author[S. J. Curran]{S. J. Curran\thanks{E-mail: sjc@phys.unsw.edu.au}\\
School of Physics, University of New South Wales, Sydney NSW 2052, Australia}

\begin{document}

\date{Accepted ---. Received ---; in original form ---}

\pagerange{\pageref{firstpage}--\pageref{lastpage}} \pubyear{2009}

\maketitle

\label{firstpage}

\begin{abstract}
We investigate the effect of two important, but oft neglected, factors
which can affect the detectability of \HI\ 21-cm absorption in Mg{\sc
  \,ii} absorption systems: The effect of line-of-sight geometry on
the coverage of the background radio flux and any possible correlation
between the 21-cm line strength and the rest frame equivalent width of
the Mg{\sc \,ii} 2796 \AA\ line, as is seen in the case of damped
Lyman-$\alpha$ absorption systems (DLAs). Regarding the former, while
the observed detection rate at small angular diameter distance ratios
($DA_{\rm abs}/DA_{\rm QSO} > 0.8$) is a near certainty ($P>0.9$), for
an unbiased sample, where either a detection or a non-detection are
equally likely, at $DA_{\rm abs}/DA_{\rm QSO} \geq 0.8$ the observed
detection rate has only a probability of $P \lapp10^{-15}$ of occuring
by chance. This $\gapp8\sigma$ significance suggests that the mix of
$DA_{\rm abs}/DA_{\rm QSO}$ values at $z_{\rm abs}\lapp1$ is
correlated with the mix of detections and non-detections at low
redshift, while the exclusively high values of the ratio ($DA_{\rm
  abs}/DA_{\rm QSO}\sim1$) at $z_{\rm abs}\gapp1$ contribute to the
low detection rates at high redshift.

In DLAs, the correlation between the 21-cm line strength
($\int\!\tau\,dv/N_{\rm HI}$) and the Mg{\sc \,ii} equivalent width
(${\rm W}_{\rm r}^{\lambda2796}$) is dominated by the velocity spread
of the 21-cm line. This has recently been shown not to hold for Mg{\sc
  \,ii} systems in general. However, we do find the significance of
the correlation to increase when the Mg{\sc \,ii} absorbers with
Mg{\sc \,i} 2852 \AA\ equivalent widths of ${\rm W}_{\rm
  r}^{\lambda2852}>0.5$ \AA\ are added to the DLA sample. This turns
out to be a sub-set of the parameter space where Mg{\sc \,ii}
absorbers and DLAs overlap and the fraction of Mg{\sc \,ii} absorbers
known to be DLAs rises to 50\% \citep{rtn05}. We therefore suggest
that the width of the 21-cm line is correlated with W$_{\rm
  r}^{\lambda2796}$ for all systems likely to be DLAs and note a
correlation between W$_{\rm r}^{\lambda2852}$ (Mg{\sc \,i}) and
$N_{\rm HI}$, which is not apparent for the singly ionised lines.  Furthermore,
the 21-cm detection rate at $DA_{\rm abs}/DA_{\rm QSO} < 0.8$ rises to
$\gapp90$\% for absorbers with ${\rm W}_{\rm
  r}^{\lambda2852}>0.5$~\AA\ and large values of $DA_{\rm abs}/DA_{\rm
  QSO}$ may explain why the absorbers which have similar values of
${\rm W}_{\rm r}^{\lambda2796}$ to the detections remain
undetected. We do, however, also find the neutral hydrogen column
densities of the non-detections to be significantly lower than those
of the detections, which could also contribute to their weak
absorption. Applying the $\int\!\tau\,dv/N_{\rm HI}$--${\rm W}_{\rm
  r}^{\lambda2796}$ correlation to yield column densities for the
Mg{\sc \,ii} absorbers in which this is unmeasured, we find no
evidence of a cosmological evolution in the neutral hydrogen column
density in the absorbers searched for in 21-cm. 
%From these estimates, at $1.1\lapp z_{\rm abs}\lapp1.6$ (where $N_{\rm HI}$ measurements are rare) we find the cosmological mass density of the neutral gas to be $\Omega_{\rm neutral~gas}\approx1\times10^{-3}$, as it is at other redshifts.

\end{abstract}

\section{Introduction}
\label{intro}

Redshifted radio absorption lines can provide an excellent probe of
the contents and nature of the early Universe, through surveys which
are not subject to the same flux and magnitude limitations suffered by
optical studies. In particular, with the \HI\ 21-cm line we can probe
the evolution of large-scale structure, as well as measuring any
putative variations in the values of the fundamental constants at
large look-back times, to at least an order of magnitude the
sensitivity provided by the best optical data. (\citealt{cdk04} and
references therein).

\begin{table}%[h]
%\centering
%\begin{minipage}{100mm}
\caption{Searches for intervening redshifted \HI~21-cm
absorption systems.
The redshift range ($z_{\rm abs}$) is given as well as the
number of detections and non-detections ($n_{\rm det}$ and $n_{\rm non}$, respectively).
\label{comp}}
\begin{tabular}{@{}l c c c c  @{}} 
\hline
%\multicolumn{5}{c}{DETECTIONS} & \multicolumn{4}{c}{NON-DETECTIONS} \\
Reference      & Type    &  $z_{\rm abs}$  & $n_{\rm det}$ & $n_{\rm non}$ \\ 
 \hline                                                               
\citet{br73} & DLA & 0.69 & 1 & 0\\
\citet{rbb+76}    & DLA & 0.52 & 1 & 0 \\
\citet{wd79} & DLA & 1.78 &  1 & 0 \\
\citet{wbj81} & DLA & 1.94 & 1 & 0 \\
\citet{bm83}$^{*}$ & sub-DLA & 0.44 & 1 & 0 \\
\citet{bw83} & Mg{\sc \,ii} & 0.37--1.94 & 1 & 16 \\
\citet{wbt+85}   & DLA  & 2.04 & 1 & 0 \\
\citet{cry93}            & Lens  & 0.69 &  1 &  0\\
\citet{cld+96} & DLA & 2.77--3.20 & 0 & 3\\
\citet{dob96} & DLA  & 3.39 &  1 &  0\\
\citet{lrj+96}           & Lens  & 0.19 &  1 & 0\\
\citet{lsb+98}& DLA & 0.22--0.31 & 2 & 0\\
\citet{cdn99}            & Lens & 0.89 & 1 &  0\\
\citet{pvtc99} & DLA & 0.63 & 0 & 1\\
\citet{ck00}$^{*}$ & DLA & 0.28--0.48 & 0 & 2\\
\citet{lan00}$^{*}$ & Mg{\sc \,ii} &  0.21--0.96& 1  & 55\\
\citet{lbs00}           & Ca{\sc \,ii} &0.09 & 1 & 0\\
\citet{lb01}$^{*}$  & Mg{\sc \,ii} &0.44 & 1& 0\\ % took out 0248+430 as in lan00 in plots?
\citet{kc01a} & DLA & 0.25--0.56 & 2 & 1 \\
\citet{kcsp01} & Mg{\sc \,ii} &0.10 & 0 & 1\\
\citet{kc02} & DLA  & 0.42--3.18 & 1 & 9\\
\citet{kb03}             & Lens & 0.76 & 1 &  0\\
\citet{dgh+04}           & ---  & 0.78 &  1 &  --\\
\citet{kse+06}           & DLA  & 2.35 &  1 &  0\\
%\citet{kcl06}$^{\dagger}$  & DLA  & 3.39 &  1 &  \\ 0201+113 detection by dob96
%\citet{gsp+06}           & Mg{\sc \,ii} & 3 & 1.17--1.37 & $0.4 - 2\times10^{18}.({T_{\rm s}}/{f})$ &    &   &  & \\
\citet{cdbw07}           & Lens  & 0.96 & 1 & -- \\
\citet{ctp+07}           & DLA  & 0.66--2.71 & 1 &2  \\
\citet{ykep07}           & DLA  & 2.29 &  1&  --\\
\citet{gsp+09a}          & Mg{\sc \,ii} & 1.10--1.45 & 9 &  26 \\                                
\citet{kpec09}$^{*}$& Mg{\sc \,ii} & 0.58--1.70 & 4 &  35\\
Zwaan et al. (in prep.)  & Mg{\sc \,ii} & $\sim0.6$  & 2 & --\\
\hline
\end{tabular}
%{DLA--damped Lyman-$\alpha$ absorption system (Mg{\sc \,ii}--DLA candidate),  Lens--gravitational lens.\\
{$^{*}$Other detections reported, but which also appear in previous papers.}
%An additional five detections reported here but these are common to \citet{gsp+09a}.}
%\end{minipage}
\end{table}
However redshifted \HI\ 21-cm absorbers are currently rare, with only
40 ``associated'' systems being detected in the hosts of radio
galaxies and quasars (see table 1 of \citealt{cww+08}) and another 40
``intervening'' systems, which lie along the sight-lines to distant
Quasi-Stellar Objects (QSOs), Table \ref{comp}. In both cases, 21-cm
absorption is predominantly detected at redshifts of $z\lapp1$, which
for the associated systems could be due to the excitation/ionisation
caused by the proximity to the active nucleus, where
optical surveys tend to select the most UV
luminous sources at high redshift (\citealt{cww+08}).%,cww08l} and references therein).

For the intervening systems, many arise in known damped Lyman-$\alpha$
absorption systems, which, at redshifts of $z_{\rm abs}\gapp1.7$,
have the Lyman-$\alpha$ line shifted into the optical band, allowing
direct measurements of the neutral hydrogen column densities ($N_{\rm
  HI}\geq2\times10^{20}$ \scm, by definition). Non-detections can be
thereby be attributed to high spin temperatures
\citep{kc02} and/or poor coverage of the background flux
\citep{cmp+03} in the high redshift systems (see Equ.~\ref{enew}, Sect. \ref{gen}).

Note, however, of the DLAs, the vast majority detected in 21-cm are
also known Mg{\sc \,ii} absorbers, these traditionally being
considered good candidates for the detection of 21-cm absorption at
$z_{\rm abs}\lapp1.7$, where the Lyman-$\alpha$ band is attenuated by the
atmosphere. Two recent surveys of Mg{\sc \,ii} systems
(at $1.10 < z_{\rm abs} < 1.45$, \citealt{gsp+09a} and $0.58 < z_{\rm
  abs} < 1.70$, \citealt{kpec09}), have found a total of 13 new 21-cm
absorbers between them, significantly increasing the number known at
$z_{\rm abs}\approx1$ (Table \ref{comp}). In these works, various
parameters (related to the equivalent widths of the singly ionised and
neutral metal lines) are discussed, although the effects of geometry
are generally ignored: \citet{cw06} attribute the high 21-cm detection rate in
DLAs identified through Mg{\sc \,ii}, cf. Lyman-$\alpha$, absorption
to the fact that the Mg{\sc \,ii} transition traces a lower redshift range ($0.2\lapp
z_{\rm abs} \lapp 2.2$, with ground-based telescopes, cf. $z_{\rm
  abs}\gapp1.7$). At redshifts of $z_{\rm abs}\gapp1.6$, the geometry
of our flat expanding Universe, ensures that foreground absorbers are
{\em always} at larger angular diameter distances than the background
QSOs, meaning that their effective coverage of the background radio
continuum is generally reduced compared to the $z_{\rm abs}\lapp1.6$
absorbers (particularly those at $z_{\rm abs}\lapp1.0$). 
%Since the Lyman-$\alpha$ line selects  higher redshifts, it stands to reason that the low 21-cm detection rate is (at least partly) due to the less effective coverage introduced by the geometry. 
In this paper, we
address this issue, investigating possible geometry effects on the
Mg{\sc \,ii} sample as a whole, as well as discussing other
possible effects on the detectability of 21-cm absorption.

\section{Factors affecting the detection of 21-cm}% absorption}
\subsection{General}
\label{gen}

The detectability of 21-cm absorption in a system lying along the
sight-line to a radio source is determined from the total
neutral hydrogen column density, $N_{\rm HI}$, via
\begin{equation}
%N_{\rm HI}=-1.823\times10^{18}.T_{\rm spin}\int\!\ln\left(1-\frac {\sigma}{f.S}\right)\,dv\,,
N_{\rm HI}=1.823\times10^{18}\,T_{\rm spin}\int\!\tau\,dv\,,
\label{enew}
\end{equation}
where $T_{\rm s}$ [K] is the mean harmonic spin temperature of the gas
and $\int\!\tau\,dv$ [\kms] is the velocity integrated optical depth of the
line. The optical depth is defined 
via $\tau\equiv-\ln\left(1-\frac{\sigma}{f\,S}\right)$, where $\sigma$
is the depth of the line (or r.m.s. noise in the case of a
non-detection) and $S$ and $f$ the flux density and covering factor of
the background continuum source, respectively.  Therefore in the
optically thin regime ($\sigma\ll f.S$), Equ. \ref{enew} reduces to
$N_{\rm HI}=1.823\times10^{18}\frac{T_{\rm spin}}{f}\int\!\frac
{\sigma}{S}\,dv$, and so with a measurement of $N_{\rm HI}$ (from the
Lyman-$\alpha$ line), the velocity integrated ``optical depth'',
$\int\!\frac {\sigma}{S}\,dv$, gives the ratio of the spin temperature
to the covering factor, $T_{\rm spin}/f$.

The importance of each of these terms to the $T_{\rm spin}/f$
degeneracy is the subject of much debate, i.e. whether the large
number of non-detections at high redshift are predominately due to
high spin temperatures \citep{kc02}, or whether the coverage of the
background emission region also plays a r\^{o}le \citep{cmp+03}.  For
the majority of the Mg{\sc \,ii} systems, which have not been observed
in the Lyman-$\alpha$ line, the situation is further complicated by
the fact that the total neutral hydrogen column density is unknown,
giving a threefold degeneracy from the integrated line
strength. Therefore determining the relative spin temperature and
covering factor contributions is currently impossible, although we can
investigate possible geometry effects, which are independent of the
assumptions used to determine these three unknowns while having a
direct bearing on the covering factor.

\subsection{Geometry effects}
\label{ge}

As stated above, \citet{cw06} suggested that for DLAs (then 17 
detections and 18 non-detections), the  21-cm detection
rate  could be influenced by geometry effects, where at
redshifts of $z_{\rm abs}\lapp1$ a foreground absorber can have a mix
of angular diameter distance ratios, $DA_{\rm abs}/DA_{\rm QSO}$,
depending upon the relative absorber and QSO redshifts, whereas at $z_{\rm
  abs}\gapp1$ the diameter distance ratio is {\em always} large
($DA_{\rm abs}/DA_{\rm QSO}\approx1$).  This is illustrated in
Fig. \ref{0.8_0.01} (top), where we show the $DA_{\rm abs}/DA_{\rm
  QSO}$ distribution for the whole Mg{\sc \,ii} absorber sample:
\begin{figure*}
\centering \includegraphics[angle=0,scale=0.85]{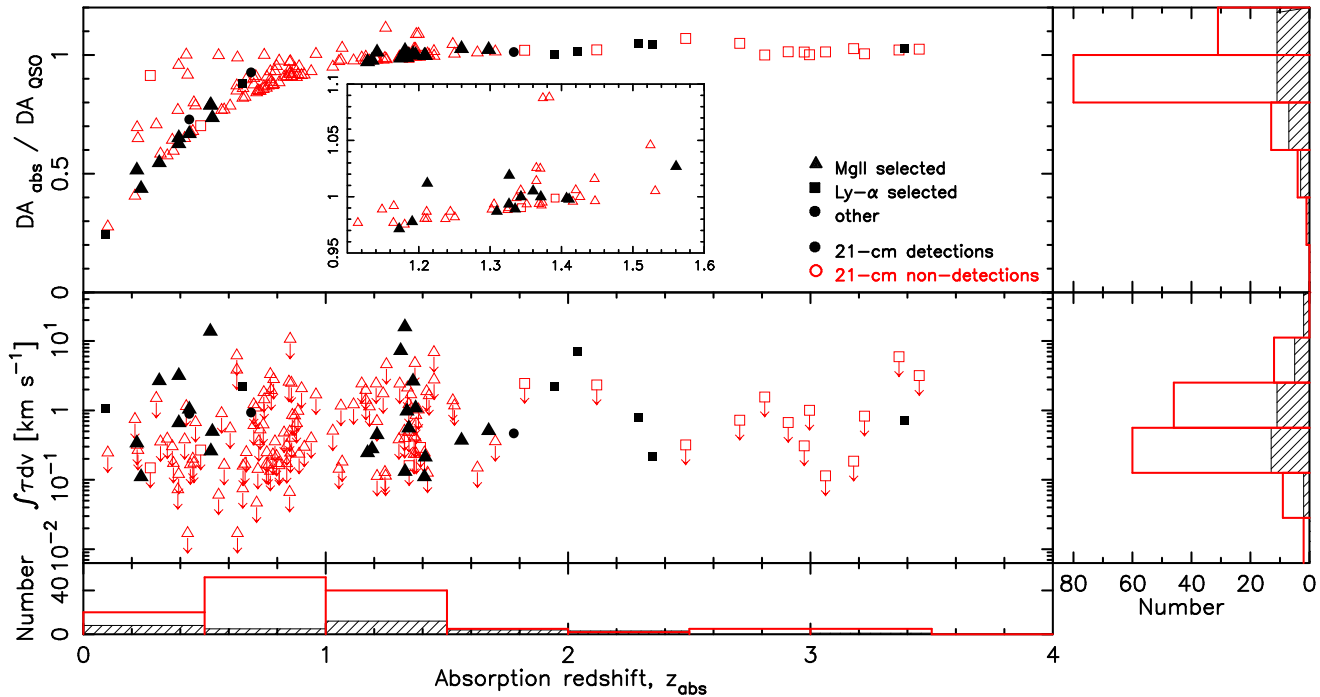}
\caption{Top: The absorber/quasar angular diameter distance ratio
  versus the absorption redshift. Note the mix of $DA_{\rm abs}/DA_{\rm QSO}$
  ratios at $z_{\rm abs}\lapp1$, whereas at $z_{\rm abs}\gapp1$ the ratio is always large.
The filled symbols/hatched histogram
  represent the 21-cm detections and the unfilled symbols/histogram
  the non-detections, with the shapes designating the transition
  through which the absorber was discovered. The inset shows the detail
  around the $z_{\rm abs} = 1.1 - 1.6$ region. Bottom: The velocity
  integrated optical depths versus the absorption redshift, where the
  arrows designate upper limits (for these the line-widths have been estimated from
W$_{\rm r}^{\lambda2796}$, see Sect. \ref{pw}). In both plots we show the DLAs plus all of the Mg{\sc
    \,ii} absorption (${\rm W}_{\rm r}^{\lambda2796} > 0$ \AA) systems searched in 21-cm.}
\label{0.8_0.01}
\end{figure*}
While the histogram for the line strengths appears to have a Gaussian
distribution (Fig.~\ref{0.8_0.01}, bottom), the
angular diameter distance ratio histogram is heavily skewed towards $DA_{\rm
  abs}/DA_{\rm QSO}=1$, where a large number of the searches for
intervening 21-cm absorption have recently occured
\citep{gsp+09a,kpec09}. It is also evident that above angular diameter
distance ratios of $DA_{\rm abs}/DA_{\rm QSO}=0.8$, that the ratio of
detections to non-detections drops drastically (shown by the upper two
histogram bars in Fig. \ref{0.8_0.01}, top)\footnote{The choice of a
  cut at $DA_{\rm abs}/DA_{\rm QSO}=0.8$ is somewhat arbitrary, but,
  as \citet{cw06}, we use this value since it is the lowest which
  gives an appreciable enough sample size in the lower bin.}.

This, however, is based on a relatively small sample at $DA_{\rm
  DLA}/DA_{\rm QSO}\leq0.8$ and so we refer to the binomial
probabilities of such a distribution occuring by chance: Assuming
that there is an equal probability of either a detection or a
non-detection occuring in either $DA_{\rm abs}/DA_{\rm QSO}$ bin, we
see that at $DA_{\rm abs}/DA_{\rm QSO}<0.8$ there is a near
certainty of obtaining the observed number detections of detections,
$P(\geq k/n)>0.9$ (Table~\ref{stats}).
However, in the $DA_{\rm abs}/DA_{\rm
    QSO}\geq0.8$ bin, the probability of the observed number of non-detections
or more occuring by chance is very small, $P(\geq k/n)\lapp10^{-15}$.
\begin{table*}
\centering
\begin{minipage}{185mm}
\caption{The statistics of the whole sample [Fig. \ref{0.8_0.01}],
  that of the absorbers with W$_{\rm r}^{\lambda2796}\geq1.0$ \AA,
  those likely to be DLAs according to \citet{rtn05}, as well as those
  likely to be DLAs by our definition [Sect. \ref{ew}]. We show the
  number of detections/total with angular diameter distance ratios of
  $DA_{\rm abs}/DA_{\rm QSO}<0.8$ and the number of
  non-detections/total with $DA_{\rm abs}/DA_{\rm QSO}\geq0.8$ with
  the binomial probability of this number or more of occuring by
  chance. ``rate'' gives the detection rate for each $DA_{\rm
    abs}/DA_{\rm QSO}$ bin.\label{stats}}
\begin{tabular}{@{}l cccccc cccccc }
\hline
Sample  &       \multicolumn{6}{c}{Lyman-\AL\ and Mg{\sc \,ii} }                          & \multicolumn{6}{c}{Mg{\sc \,ii}  only}\\
         & \multicolumn{3}{c}{\sc detections} &  \multicolumn{2}{c}{\sc non-detections} & \multicolumn{3}{c}{\sc detections} &  \multicolumn{3}{c}{\sc non-detections}\\
                & $<0.8$ & $P(\geq k/n)$ & rate & $\geq0.8$ & $P(\geq k/n)$ & rate & $<0.8$  & $P(\geq k/n)$ & rate & $\geq0.8$   & $P(\geq k/n)$ & rate  \\
\hline
%\multicolumn{13}{c}{\citet{lan00} and \citet{gsp+09a}}\\
%\hline
%Whole       &   11/28       &   0.91  &    39\%   &  80/98   & $8.30\times10^{-11}$ &  18\%   &  9/25  &   0.95  &  36\%  &  64/74 & $4.45\times10^{-11}$ &  14\% \\
%W$_{\rm r}^{\lambda2796}\geq1.0$ \AA & 11/13 & 0.011  & 85\%   & 59/75  & $3.06\times10^{-7}$&   21\%   &  9/10    & 0.011  &  90\%  &  43/51& $3.43\times10^{-7}$      &   16\%   \\
%DLAs \citep{rtn05}   & 8/9   &0.020  &   89\%   &  24/37  &  0.049  &   35\%    & 7/7   &  0.0078 &  100\% &  13/19&   0.084&    32\%\\
%\hline
%\multicolumn{13}{c}{\citet{lan00,gsp+09a} and \citet{kpec09}}\\
%\hline
Whole       &   11/29       &   0.93  &    38\%   &  111/133   & $8.55\times10^{-16}$ &  17\%   &  9/26  &   0.96  &  35\%  &  95/109 & $2.88\times10^{-16}$ &  13\% \\
W$_{\rm r}^{\lambda2796}\geq1.0$ \AA & 11/14 & 0.029  & 79\%   & 88/106  & $1.49\times10^{-12}$&   17\%   &  9/11    & 0.033  &  82\%  &  72/82& $5.11\times10^{-13}$ & 12\% \\
%% NOT USING W < 3.3 UPPER LIMIT FOR THIS NEXT ONE
\citet{rtn05}   & 8/9   &0.020  &   89\%   &  44/60  & $1.97\times10^{-4}$   &  27\%    & 7/7   & 0.0078 &  100\% &  33/42& $1.36\times10^{-4}$  &  21\%\\

${\rm W}_{\rm r}^{\lambda2852}>0.5$ \AA & 9/10   &0.011  &   90\%   &  44/64  & 0.0020   &  31\%    & 7/7   & 0.0078 &  100\% &  33/46& 0.0023  &  28\%\\
%from 2-distance-z_our.c
% run 2-distance-z_rtn_fig11.c to get as Fig. 11 of rtn05 - just about identical to plain old rtn05, row 3
\hline
\end{tabular}
\end{minipage}
\end{table*}

For the ${\rm W}_{\rm r}^{\lambda2796} > 0.6$ \AA\ Mg{\sc \,ii}
absorbers with $1 < {\rm W}_{\rm r}^{\lambda2796}/{\rm W}_{\rm
  r}^{\lambda2600}<2$ {\sc and} ${\rm W}_{\rm r}^{\lambda2852}>0.1$
\AA, i.e. those in which $\approx40\%$ are known to be DLAs
\citep{rtn05}, we  obtain a similar distribution at $DA_{\rm
  abs}/DA_{\rm QSO}\geq0.8$, although the
probabilities are much higher (row 3 of Table \ref{stats}), due to the
smaller sample. However, a probability of $P(\geq k/n) = 1.36
\times10^{-4}$ is still significant at $3.81\sigma$, assuming Gaussian
statistics, compared with $P(\geq k/n)=0.026 \Rightarrow2.23\sigma$
for the confirmed DLAs only \citep{cw06}, where there were 16
non-detections out of a sample of 22 at $DA_{\rm abs}/DA_{\rm
  QSO}\geq0.8$\footnote{Combined with the $\geq11/13$ detections at
  $DA_{\rm abs}/DA_{\rm QSO}<0.8$, this gives an overall probability
  of $0.00029 (\Rightarrow3.63\sigma)$ for the whole distribution.}.

From the bottom  panel of Fig. \ref{0.8_0.01}, it is
apparent that the 21-cm surveys cover a wide range of sensitivities
and with no knowledge of the total neutral hydrogen column densities
in many of these, it is not possible to normalise out the
observational biases (Sect. \ref{gen}). However, the  histogram in
Fig. \ref{0.8_0.01} shows that, on the whole, the
non-detections have been searched as deeply as the detections and,
given that the observed distribution 
may be driven by other effects\footnote{Such as different $T_{\rm
    spin}/f$ ratios (see \citealt{ctp+07}) or column densities (see
  Sect. \ref{rdng}).}, all else being equal, the angular diameter
distance ratio does appear to be correlated with the detection
rate. 

To recap, for all of the {Mg{\sc \,ii} absorption systems the
  observed detection rate at low angular diameter distance ratios has
  a high probability of occuring by chance, while that at high ratios
  is extremely unlikely and, when placing conditions on the
  sample, by selecting those which could be DLAs, there is an
  extremely high detection rate at low angular diameter distance
  ratios, while the rate remains low at high ratios.
Since, on the basis that a given absorption cross-section will 
cover a given emission region less effectively when these are
at similar angular diameter distances, it stands to reason that
the observed distribution is, at least in part, driven by the
line-of-sight geometry.

\subsection{Mg{\sc \,ii} equivalent width}
\subsubsection{21-cm line width}
\label{ew}

While the above work demonstrates that the detection rate appears to
be dependent upon the angular diameter distance ratio, the strength of
the 21-cm absorption (normalised by the neutral hydrogen column
density, $\int\!\tau\,dv/N_{\rm HI}$) may itself be related to the
Mg{\sc \,ii} equivalent width, as shown to apply to the confirmed DLAs
\citep{ctp+07}. However, no such correlation was found between
$\tau$ and ${\rm W}_{\rm r}^{\lambda2796}$, with only a weak trend
existing between $\int\!\tau\,dv$ and ${\rm W}_{\rm
  r}^{\lambda2796}$. The correlation therefore appears to be dominated
by the velocity spread of the 21-cm profile (both FWHM and total
velocity spread), ``significant'' at $1.80\sigma$, which rises to
$2.30\sigma$ when the outlier 1622+238 is removed\footnote{\label{foot3}\citet{ctp+07} previously reported a
  significance of $2.21\sigma$ ($2.84\sigma$ without 1622+238) for the
  FWHM--${\rm W}_{\rm r}^{\lambda2796}$ correlation, but here we
  exclude 0248+430 (at ${\rm W}_{\rm r}^{\lambda2796}=1.86$ \AA,
  FWHM$\,=19$ \kms) which, while generally being regarded as a DLA, has
  an unknown neutral hydrogen column density. Also, we have included
  2003--025 \citep{kpec09}, which at ${\rm W}_{\rm
    r}^{\lambda2796}=0.74$ \AA, FWHM$\,=40$ \kms, is somewhat of an
  outlier.}: With a
FWHM$\,=235$ \kms\ and an impact parameter of $\approx75$ kpc
(\citealt{ctm+07} and references therein), this is a most unusual DLA
in which we believe the 21-cm profile width is dominated by
large-scale dynamics.

\citet{gsp+09a,kpec09} report no correlation between the FWHM and ${\rm W}_{\rm r}^{\lambda2796}$
for the Mg{\sc \,ii} absorbers and in 
\begin{figure}
\centering \includegraphics[angle=270,scale=0.72]{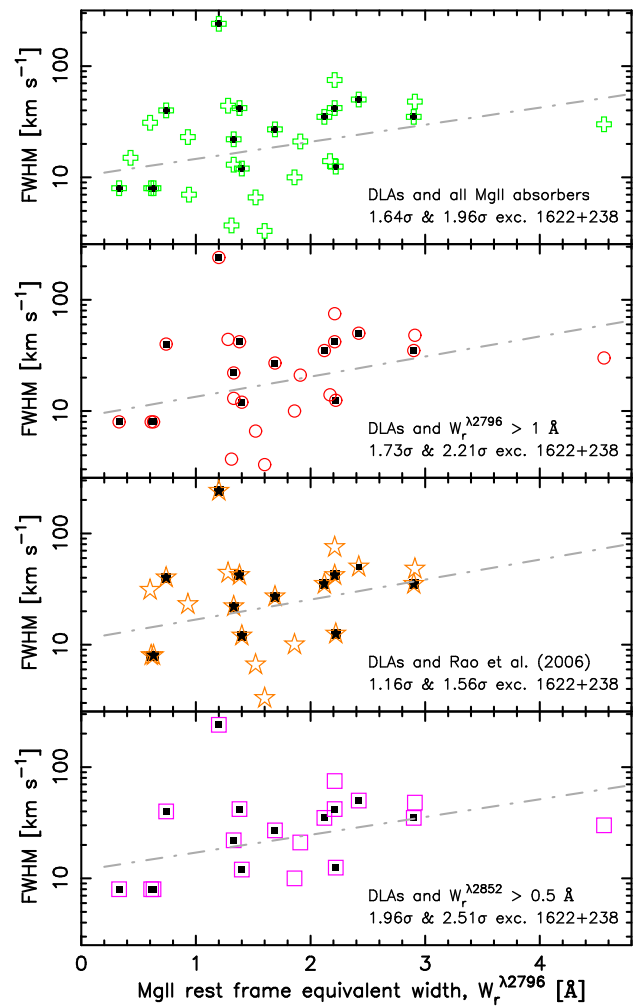}
\caption{The full width half maximum of the 21-cm absorption profile
  versus the Mg{\sc \,ii} 2796 \AA\ equivalent width for the DLAs and
  Mg{\sc \,ii} absorption systems. All of the panels show the
  confirmed DLAs (small black squares) overlain with the unfilled symbols,
  which show various Mg{\sc \,ii} absorber sub-samples -- all Mg{\sc
    \,ii} absorbers detected in 21-cm, those with 2796 \AA\ equivalent
  widths of W$_{\rm r}^{\lambda2796}>1$ \AA, those which could be DLAs
  according to \citet{rtn05} [$1 < {\rm W}_{\rm r}^{\lambda2796}/{\rm
      W}_{\rm r}^{\lambda2600}<2$ {\sc and} ${\rm W}_{\rm
      r}^{\lambda2852}>0.1$ \AA, where $0.6 \leq {\rm W}_{\rm
      r}^{\lambda2796} < 3.3$ \AA] and finally those with just ${\rm
    W}_{\rm r}^{\lambda2852}>0.5$~\AA.  In each panel we give the
  significance of the correlation for the DLAs plus Mg{\sc \,ii}
  absorbers with and without the outlier 1622+238, with the line
  showing the least-squares fit to the latter.}
\label{fwhm_EW}
\end{figure}
Fig. \ref{fwhm_EW} we show the confirmed DLAs (where $N_{\rm
  HI}\geq2\times10^{20}$ \scm) together with various sub-sets of
Mg{\sc \,ii} absorbers detected in 21-cm and do find that the addition
of the Mg{\sc \,ii} absorbers degrades the correlation (top
panel). This regains a similar significance to that of the DLAs only
\citep{ctp+07} when only the Mg{\sc \,ii} absorbers with ${\rm W}_{\rm
  r}^{\lambda2796} >1$ \AA\ are added (second panel) and degrades
once more with the condition of \citet{rtn05} applied to the Mg{\sc
  \,ii} absorbers which overlap the same ${\rm W}_{\rm
  r}^{\lambda2796}/{\rm W}_{\rm r}^{\lambda2600}$---$\,{\rm W}_{\rm
  r}^{\lambda2852}$ space as the confirmed DLAs
(third panel).

Through various trials, we only find a significant increase in the
correlation for the Mg{\sc \,ii} absorbers with ${\rm W}_{\rm
  r}^{\lambda2852}>0.5$ \AA\ added to the confirmed DLAs (bottom
panel). Although found empirically, our condition actually selects a
sub-set of the range specified by \citet{rtn05}, where $\approx40$\%
of the Mg{\sc \,ii} absorbers are DLAs. At ${\rm W}_{\rm
  r}^{\lambda2852}>0.5$ \AA, 16 of the 30 absorbers are DLAs (50\%),
cf. 16 out of 49 (30\%) at $1 < {\rm W}_{\rm r}^{\lambda2796}/{\rm
  W}_{\rm r}^{\lambda2600}<2$ {\sc and} $0.1< {\rm W}_{\rm
  r}^{\lambda2852}<0.5$ \AA\ (figure 11 of \citealt{rtn05}). Above
${\rm W}_{\rm r}^{\lambda2852}\approx1$ \AA, all of the Mg{\sc \,ii}
absorbers are DLAs, although being a sample of only four severely
restricts the significance of this\footnote{For example, applying this
  condition to the 21-cm absorbers gives only three Mg{\sc \,ii}
  systems in addition to the DLAs (cf. the already low number of five
  when applying ${\rm W}_{\rm r}^{\lambda2852}>0.5$ \AA,
  Fig. \ref{fwhm_EW}). These three do increase the significance of the
  correlation slightly to $1.82\sigma$ ($2.36\sigma$ without 1622+238).}.

Note that, although our ${\rm W}_{\rm r}^{\lambda2852}>0.5$
\AA\ selection is a sub-set of the DLA range of \citet{rtn05}, our
FWHM--${\rm W}_{\rm r}^{\lambda2796}$ correlation is significantly
higher than when applying their condition. This can be attributed to
the inclusion of the end point at ${\rm W}_{\rm r}^{\lambda2796}=4.56$
\AA\ and FWHM$\,=30$ \kms\ (towards J0850+5159,
\citealt{gsp+09a})\footnote{Excluded in the $0.6 \leq {\rm W}_{\rm
    r}^{\lambda2796} < 3.3$ \AA\ \citet{rtn05} sample.}, as well as a
tighter selection which introduces some differences in the sample at
${\rm W}_{\rm r}^{\lambda2796} \lapp2$ \AA\ (the third cf. the bottom
panel of Fig. \ref{fwhm_EW}).

\subsubsection{21-cm line strength}
\label{pw}

Although the normalised 21-cm line strength may be correlated with the
Mg{\sc \,ii} 2796 \AA\ equivalent width for the confirmed DLAs, \citet{cw06} noted that the 21-cm
non-detections span a similar range of equivalent widths. Therefore, while
${\rm W}_{\rm r}^{\lambda2796}$ may be a diagnostic of the 21-cm line strength
($\int\!\tau\,dv/N_{\rm HI}\propto f/T_{\rm spin}$), it cannot predict whether 
or not 21-cm will be detected. One possible explanation is that
the non-detections are disadvantaged through geometry effects, as 
discussed in Sect. \ref{ge}.

\begin{figure*}
\centering \includegraphics[angle=270,scale=0.75]{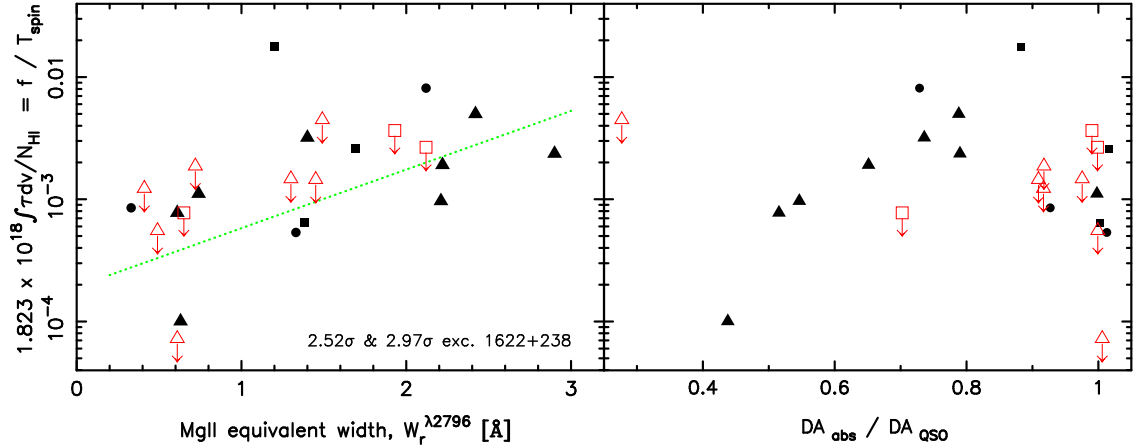}
\caption{The 21-cm line strength  versus the rest frame
  equivalent width of the Mg{\sc \,ii} 2796 \AA\ line (left) and the
  angular diameter distance ratio (right) for the DLAs searched in
  21-absorption. The line in the left panel shows the least-squares
  fit to all of the points. The symbols are as per Fig. \ref{0.8_0.01}.}
\label{strength_W_ratio}
\end{figure*}
In Fig. \ref{strength_W_ratio} we show the line strength against the
equivalent width together with the angular diameter distance ratios.
From this, we see that the non-detections are indeed disadvantaged,
with most of these being at angular diameter distance ratios of
$\gapp0.9$, whereas the detections have a range of ratios, with the
large spiral galaxies located at the high end with the strongest line
strengths (see figure 6 of \citealt{ctp+07}). This is strong evidence
that the geometry is, once again, a dominant effect, although applying
a survival analysis to the non-detections\footnote{Via the {\sc asurv}
  package \citep{ifn86}.}, raises the significance of the correlation
of the whole sample quite dramatically: $2.52\sigma$ ($2.97\sigma$
without 1622+238), cf. $1.59\sigma$ ($2.07\sigma$ without
1622+238)\footnote{Updated from \citet{ctp+07}, as per the changes
  described in footnote \ref{foot3}.} for the detections only
\citep{ctp+07}. Hence, if the 21-cm line strength and Mg{\sc \,ii}
equivalent width are related, this suggests that many of the
non-detections may only require slightly deeper searches.

The values used to derive the limits for the 21-cm non-detections
(Fig. \ref{strength_W_ratio}) will of course be biased by the use of the
correlation for the detections to estimate the FWHM of the
non-detections. This gives FWHM\,$\approx13\,{\rm W}_{\rm
  r}^{\lambda2796}$ (figure 6 of \citealt{ctp+07}), and, where the
Mg{\sc \,ii} equivalent widths are not available (generally at $z_{\rm
  abs}\gapp2.2$), we estimate these from the metallicity via ${\rm
  W}_{\rm r}^{\lambda2796}\approx2.0\,{\rm [M/H]} + 4.0$ (figure 7 of
\citealt{ctp+07}). When none are available, we  apply the
average 20 \kms\ of the DLAs detected in 21-cm absorption
\citep{cmp+03}. Here, the 21-cm detections span a FWHM of 8 to 50 \kms\ (or 235
\kms\ including 1622+238), giving an average value of 26 \kms\ and
applying the methods above gives 5 to 28 \kms\ for the non-detections, with an
average value of 15 \kms. If instead we just assume the average value of the detections
as the FWHM of each non-detection, each of these moves these further up the ordinate
in Fig. \ref{strength_W_ratio}, while reducing the significance of the
correlation to $2.07\sigma$ ($2.56\sigma$ without 1622+238).

The fact that, for the DLAs, the normalised line strength exhibits the
strongest correlation with equivalent width (followed by the FWHM then
$\int\!\tau\,dv$, \citealt{ctp+07}), suggests that these parameters are inseparable,
giving ${\rm W}_{\rm r}^{\lambda2796}\propto f/T_{\rm spin}$, where
$f$ is generally lower for the non-detections as a result of the
higher angular diameter distance ratios. In the absence of measured neutral
hydrogen column densities, this means that we cannot use the majority
of the Mg{\sc \,ii} absorbers to verify this. Conversely, the correlation
may provide a method with which to estimate these column densities, one
use of which we now explore.

\section{Redshift distribution of the neutral gas}
\label{rdng}

Using the fit from the 21-cm line strength--Mg{\sc \,ii} 2796
\AA\ equivalent width correlation (Fig. \ref{strength_W_ratio}), we
can estimate total neutral hydrogen column densities for the Mg{\sc
  \,ii} absorbers for which these are unavailable, which we show in 
Fig. \ref{3-N}.
\begin{figure}
\centering \includegraphics[angle=270,scale=0.75]{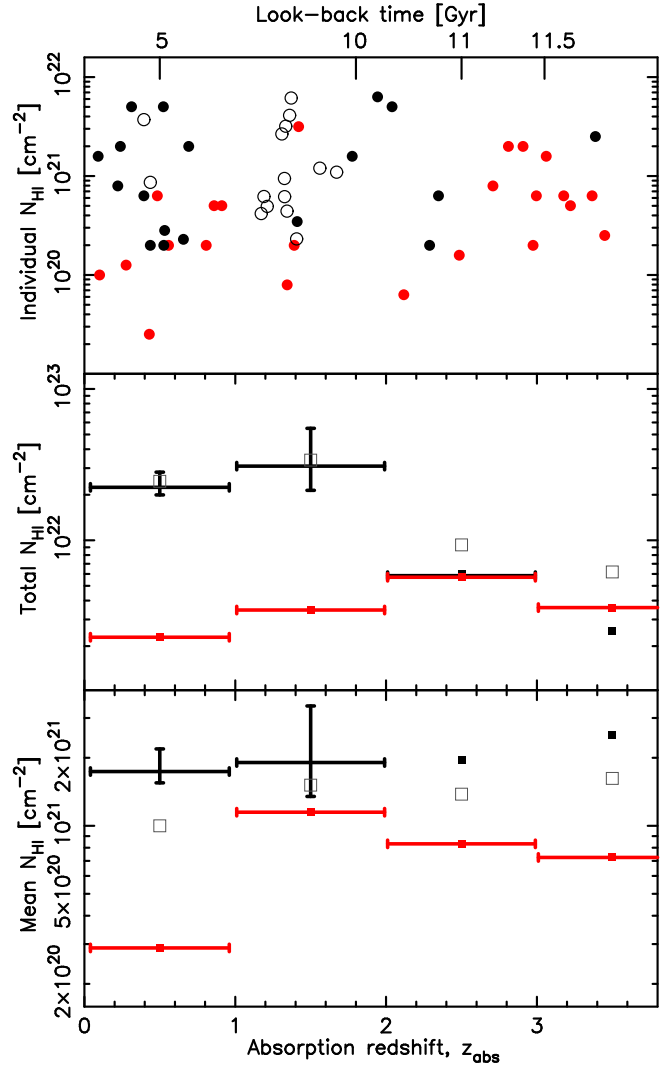}
\caption{The column density--redshift distribution of the Mg{\sc \,ii}
  absorbers searched for in 21-cm, where the black
  markers denote the detections and the coloured markers the
  non-detections Top: The individual column densities, where the
  filled markers show the actual column density measurements and the
  unfilled markers show the column density calculated from the 21-cm
  line strength as per the fit in Fig. \ref{strength_W_ratio}.  We do
  not include the limits determined from the 21-cm non-detections for
  which there are no measurements of $N_{\rm HI}$. Middle: The total
  values in $\Delta z =1$ redshift bins.  Bottom: The mean $N_{\rm
    HI}$ per absorber within each bin . The error bars on
  the ordinate show the $1\sigma$ uncertainty to the fit in
  Fig. \ref{strength_W_ratio} and the unfilled squares show
  the detected and non-detected values combined.}
\label{3-N}
\end{figure}
What is immediately clear is that the 21-cm non-detections have
systematically lower column densities than the absorbers
detected in 21-cm, particularly at low redshift ($0 < z_{\rm abs} <
1$, where many do have a measured $N_{\rm HI}$)\footnote{Although the
  column density is normalised out in Fig. \ref{strength_W_ratio},
  suggesting that, apart from the angular diameter distance ratios,
  these have been searched sufficiently deeply.}. Although there is
considerable overlap in the column densities between the detections
and the non-detections, the lower values combined with the geometry
issues addressed above, could account for the paucity of strong 21-cm
absorption in these objects. \citet{cmp+03} previously noted a
correlation between the velocity integrated optical depth and the
column density for the confirmed DLAs, which is found to be
significant at $3.08\sigma$ when the non-detections are also included
\citep{ctd+09}.

In the middle panel of Fig. \ref{3-N} we show the total column density
for all of the 21-cm searched absorbers, which have a
measured/estimated value of this.  It is apparent that, after a
possible peak in the $1 < z_{\rm abs} < 2$ bin, $N_{\rm HI}$ decreases
with redshift. However, this will be heavily influenced by the
relative paucity of high redshift searches, due to the availability of
specific radio bands in conjunction with a severe radio interference
environment, which can hamper searches at these frequencies.  This
effect, in addition to the geometry effects discussed above, could explain the steep
decrease in the number of 21-cm absorbers with redshift
\citep{gsp+09a}, which runs contrary to the increase in the number of
DLAs \citep{rtn05}.

Therefore in the bottom panel we show the mean column density per bin
which, although limited by the very few points in the high redshift bins,
shows no evidence of an evolution of $N_{\rm HI}$.
Using these combined averages ($\overline{N_{\rm HI}}$), we may calculate the cosmological mass density
of the neutral gas in the Universe as a function of redshift, via:
%\begin{equation}
\[
\Omega_{\rm neutral~gas} = \frac{\mu\,m_{_{\rm H}}\,H_{0}}{c\,\rho_{\rm crit}}\,n_{\rm DLA}\,\overline{N_{\rm HI}}\frac{1}{(z+1)^2}\frac{H_{\rm z}}{H_{0}},
\]
%\end{equation}
where $\mu = 1.3$ is a correction for the 75\% hydrogen composition,
$m_{_{\rm H}}$ is the mass of the hydrogen atom, $H_{0}=71$~\kms\ is
the present Hubble parameter, $c$ the speed of light, $\rho_{\rm
  crit} \equiv 3\,H_{0}^2/8\,\pi\,G$ is the critical mass
density of the Universe\footnote{Where $G$ is the gravitational constant.} and 
%\begin{equation}
\[
\frac{H_{\rm z}}{H_{0}} = \sqrt{\Omega_{\rm matter}\,(z+1)^3 + (1-\Omega_{\rm matter} - \Omega_{\Lambda})\,(z+1)^2 + \Omega_{\Lambda}},
%\end{equation}
\]
where $H_{\rm z}$ is the Hubble parameter at redshift $z$ and we use $\Omega_{\rm matter}=0.27$ and
$\Omega_{\Lambda}=0.73$. 
%\footnote{Our values are in excellent agreement with \citet{rtn05}: $\overline{N_{\rm HI}}= 1.26^{+0.33}_{-0.14} \times10^{21}$ cf.  $1.26\pm0.36\times10^{21}$ \scm\ at $0.11 < z_{\rm abs} < 0.9$ and $\overline{N_{\rm HI}}= 1.49^{+1.36}_{-0.50} \times10^{21}$ cf.  $1.26\pm0.36\times10^{21}$ \scm\ at $0.90 <z_{\rm abs} < 1.65$.}% but most of meaured Ns will be from rtn05 
From the redshift number density of DLAs, $n_{\rm DLA}$
\citep{ph04,rtn05} and deriving the mean column densities according
to how the $n_{\rm DLA}$ values are binned in redshift, we obtain the
cosmological mass density distribution shown in Fig. \ref{omega-z}.
\begin{figure}
\centering \includegraphics[angle=270,scale=0.75]{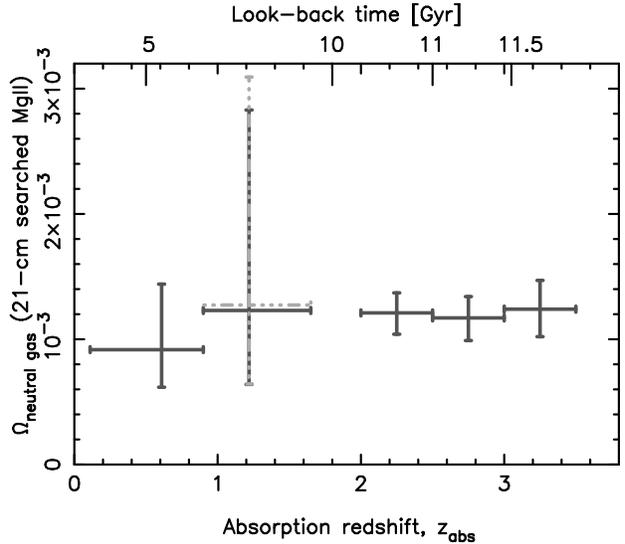}
\caption{The cosmological mass density versus redshift for the Mg{\sc \,ii}
  absorbers searched for in 21-cm. The error bars show the combined
uncertainties from $\overline{N_{\rm HI}}$ (above) and $n_{\rm DLA}$ in each redshift bin \citep{ph04,rtn05}.
The dashed--dotted bars in the $0.90 <z_{\rm abs} \leq 1.65$ bin show the result if the 
${\rm W}_{\rm r}^{\lambda2796}>0.5$ \AA\ condition of \citet{kpec09} is applied.}
\label{omega-z}
\end{figure}

From this we see a near constant $\Omega_{\rm
  neutral~gas}\approx1\times10^{-3}$, as previously noted by
\citet{rt00,ph04,rtn05}, where the cosmological mass density of the
neutral gas remains unchanged up to redshifts of $z_{\rm
  abs}\approx4$. This is not surprising as our Mg{\sc \,ii}/DLA
sample\footnote{Note that while all of the sample is not necessarily
  comprised of DLAs, from Fig. \ref{3-N} it is apparent that at least
  all of the 21-cm detections have $N_{\rm HI}\geq2\times10^{20}$
  \scm, including the column densities estimated from the 21-cm
  detections (unfilled markers). Although some of the 21-cm
  non-detections have $N_{\rm HI}\lapp10^{20}$ \scm, from the bottom
  panel of Fig. \ref{3-N} the mean value exceeds the defining DLA
  column density in all redshift bins.}  is comprised of sub-sets of
the aforementioned samples. The fact that our value at $1.1\lapp
z_{\rm abs}\lapp1.6$ is consistent with that of \citet{rtn05},
provides a check on 21-cm line strength--Mg{\sc \,ii} 2796
\AA\ equivalent width correlation
(Fig. \ref{strength_W_ratio}). However, it is nearly double the value
of $\Omega_{\rm neutral~gas}$ calculated at this redshift by
\citet{kpec09}\footnote{Even when applying their ${\rm W}_{\rm
    r}^{\lambda2796}>0.5$ \AA\ criterion (Fig. \ref{omega-z}).},
although this work assumes an average \HI\ column density at these
redshifts, whereas our value is estimated using the
%21-cm line strength--Mg{\sc \,ii} 2796 \AA\ equivalent width
correlation.

\section{Discussion and Summary}

Following the recent surveys of Mg{\sc \,ii} absorbers at $0.58 <
z_{\rm abs} < 1.70$ \citep{gsp+09a,kpec09}, there has been a large increase
in the number of intervening 21-cm absorption systems detected at
these redshifts.  Both these works discuss various reasons regarding
the detection of 21-cm absorption in Mg{\sc \,ii} systems, but these
focus upon the various equivalent widths of the singly ionised and
neutral metal lines, without considering the possible line-of-sight
geometry effects, although these have been found to bias the detection rate for confirmed
damped Lyman-$\alpha$ absorption systems \citep{cw06}. 
Considering this effect, we find for the  Mg{\sc \,ii} absorbers:
\begin{enumerate}
  \item For the sample as a whole (or limiting this to strong Mg{\sc
    \,ii} absorbers, e.g. ${\rm W}_{\rm r}^{\lambda2796} > 1$ \AA), the mix of
    detections at low redshifts is what would be expected purely from chance,
    although the high non-detection rate at $z_{\rm abs}\gapp1$ is highly
    unlikely.

    \item When restrictions are applied to the sample, so that a large
      fraction of the Mg{\sc \,ii} absorbers are expected to consist of DLAs
      (according to \citealt{rtn05}), detection rates of $\approx100$\% are
      reached at low redshift, while the low detection probabilities at  $z_{\rm abs}\gapp1$
      remain significant.
\end{enumerate}
Although other factors may contribute (such as lower column densities
for the 21-cm non-detections, Sect. \ref{rdng}), which may be
unmeasurable (i.e. $T_{\rm spin}/f$ in the absence of $N_{\rm HI}$),
it is apparent that the likelihood of detecting 21-cm absorption is
correlated with the angular diameter distance ratio for all
intervening absorption systems, where the absorbers at low redshift
are subject to a range of $DA_{\rm abs}/DA_{\rm QSO}$ ratios and thus
exhibit a mix of detections and non-detections, whereas those at
$z_{\rm abs}\gapp1$ all have $DA_{\rm abs}/DA_{\rm QSO}\sim1$, while
tending to be non-detections.

Since a given absorber will occult the background flux much more effectively at
a lower angular diameter distance ratio, this suggests a strong
contribution by the covering factor, which is independent from the
measured 21-cm line strength, from which the spin temperature/covering
factor degeneracy cannot be broken (Sect.~\ref{gen}): In some cases the covering factor
has been estimated as the ratio of the compact unresolved component's
flux to the total radio flux \citep{bw83,klm+09}. However, even if
high resolution radio images at the redshifted 21-cm frequencies are
available, this method gives no information on the extent of the
absorber (or how well it covers the emission). Furthermore, the high
angular resolution images are of the continuum only and so do not give
any information about the depth of the line when the extended
continuum emission is resolved out. By using the angular diameter distance
ratios, we completely circumvent these issues, although we can
only discuss generalities which cannot determine values of  $T_{\rm spin}$
or $f$ for individual systems. What we do find, however, from our statistically
large sample (a total of 162 absorption systems), is that 21-cm detection
rates are much lower when the absorber and background QSO are at similar
angular diameter distances.

Saying this, the recent surveys \citep{gsp+09a,kpec09} yield 13 new
detections between them at $z_{\rm abs}\sim1$, where the angular
diameter distance ratios are high ($DA_{\rm abs}/DA_{\rm
  QSO}\gapp0.8$, Fig. \ref{0.8_0.01}).  However, each of these
surveys also yields a large number of non-detections, with both
exhibiting a 25\% detection rate\footnote{The binomial probability of
  26 or more non-detections out of a sample of 35 is $P(\geq
  k/n)=0.0030$ (\citealt{gsp+09a}, where at $1.10 < z_{\rm abs} <
  1.45$ all of the systems have $DA_{\rm abs}/DA_{\rm
    QSO}\sim1$).}. This is about double the rate for the whole $DA_{\rm
  DLA}/DA_{\rm QSO}>0.8$ sample, but very close to that when
restrictions are applied to include those Mg{\sc \,ii} absorbers which
may also be DLAs (Table \ref{stats}).  This is not surprising since
the \citet{kpec09} survey selects only those with ${\rm W}_{\rm
  r}^{\lambda2796} > 0.5$ \AA\ (the minimum equivalent width of a DLA
in the \citealt{rtn05} sample -- see their figure 2) and
\citet{gsp+09a} select those with ${\rm W}_{\rm r}^{\lambda2796} > 1$
\AA, restricting this further. 

Another possible factor affecting the detection of 21-cm absorption
could be the correlation between the 21-cm line strength
($\int\!\tau\,dv/N_{\rm HI}\propto f/T_{\rm spin}$) and the Mg{\sc
  \,ii} 2796 \AA\ equivalent width (${\rm W}_{\rm r}^{\lambda2796}$,
\citealt{cw06}).  However, the same range of equivalent widths is also
spanned by the non-detections, making this a questionable diagnostic
with which to find 21-cm absorption systems. We do find, however, that
eight of the ten non-detections have $DA_{\rm abs}/DA_{\rm QSO}>0.9$,
whereas the 13 detections are much more spread out, spanning $0.4 <
DA_{\rm abs}/DA_{\rm QSO} < 1$, which suggests that geometry effects
may again be responsible for the detection rate.  If the
$\int\!\tau\,dv/N_{\rm HI}\propto f/T_{\rm spin}$--${\rm W}_{\rm
  r}^{\lambda2796}$ correlation holds up, the fact that the
non-detections increase the significance, suggests that these
absorbers may be close to 21-cm detection. In any case, we suggest
that ${\rm W}_{\rm r}^{\lambda2796}$ in conjunction with the value of
$DA_{\rm abs}/DA_{\rm QSO}$ may provide a diagnostic with which to
find 21-cm absorption.

As discussed by \citet{ctp+07}, the strongest contributer to this
correlation appears to be the 21-cm profile width, which is not surprising as at W$_{\rm
  r}^{\lambda2796}\gapp0.3$~\AA, the range discussed here, the Mg{\sc
  \,ii} profile is dominated by velocity structure
\citep{ell06}. Including all of the Mg{\sc \,ii} absorbers degrades the
correlation (as found by \citealt{gsp+09a,kpec09}), although through
the testing of many different equivalent widths and their ratios
(e.g. ${\rm W}_{\rm r}^{\lambda2796}$, ${\rm W}_{\rm r}^{\lambda2852}$
\& ${\rm W}_{\rm r}^{\lambda2796}/{\rm W}_{\rm r}^{\lambda2600}$), we
find that the correlation increases for the Mg{\sc \,ii} absorbers for
which the Mg{\sc \,i} 2852 \AA\ equivalent width is ${\rm W}_{\rm
  r}^{\lambda2852}>0.5$ \AA. At 13.60 eV, \HI\ has a similar
ionisation potential to the singly ionised metal species (15.04 for
Mg{\sc \,ii} \& 16.18 eV for Fe{\sc \,ii}), although it is not
surprising that the cooler component (i.e. the 21-cm) is better traced
by the neutral metal species (i.e. Mg{\sc \,i}, which has a potential
of 7.65 eV).  Furthermore, the fact that ${\rm W}_{\rm
  r}^{\lambda2852}>0.5$ \AA\ is a sub-set of the ${\rm W}_{\rm
  r}^{\lambda2796}/{\rm W}_{\rm r}^{\lambda2600}$---$\,{\rm W}_{\rm
  r}^{\lambda2852}$ space which contains all of the DLAs of
\citet{rtn05}, suggests that the ${\rm W}_{\rm
  r}^{\lambda2852}>0.5$ \AA\ selection is (mostly) adding further DLAs to
the confirmed DLA sample \citep{ctp+07}. That is, in general, the 21-cm
profile width is correlated with W$_{\rm r}^{\lambda2796}$ when $N_{\rm
  HI}\geq2\times10^{20}$ \scm.

Investigating this further, in Fig. \ref{pdlc} we show the total
neutral hydrogen column density versus the Mg{\sc \,i} 2852
\AA\ equivalent width for the $z_{\rm abs} < 1.7$ absorbers for which both
measurements are available.
\begin{figure}
\centering \includegraphics[angle=270,scale=0.75]{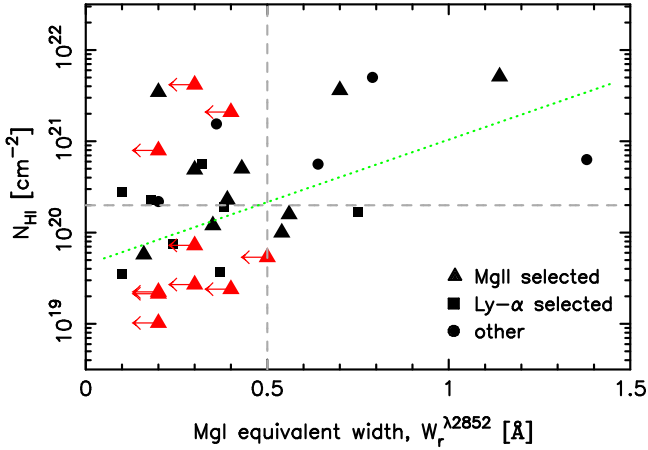}
\caption{The total neutral hydrogen column density versus the Mg{\sc
    \,i} 2852~\AA\ equivalent width for  $z_{\rm abs} < 1.7$ absorbers \citep{pdlc04}.
The coloured symbols/arrows designate upper limits to ${\rm W}_{\rm r}^{\lambda2852}$.
The horizontal dashed line shows $N_{\rm HI}=2\times10^{20}$ \scm, above which the
absorbers are considered to be DLAs and the dotted line shows the least-squares
fit to all of the points (taking account of the limits via {\sc asurv}). }
\label{pdlc}
\end{figure}
Although there is no clear partitioning, all (in an admittedly
small sample) of those with ${\rm W}_{\rm r}^{\lambda2852}>0.5$
\AA\ have $N_{\rm HI}\gapp10^{20}$ \scm, and unlike
\citet{pdlc04}, who find no significant correlation between metal line
equivalent widths (${\rm W}_{\rm r}^{\lambda2796}$ \& ${\rm W}_{\rm
  r}^{\lambda2600}$) and $N_{\rm HI}$, we find ${\rm W}_{\rm r}^{\lambda2852}$
to correlate with $\log_{10}N_{\rm HI}$ at a $2.83\sigma$ significance.

\begin{figure}
%\centering \includegraphics[angle=270,scale=0.75]{2852_ratio.ps}
\centering \includegraphics[angle=270,scale=0.75]{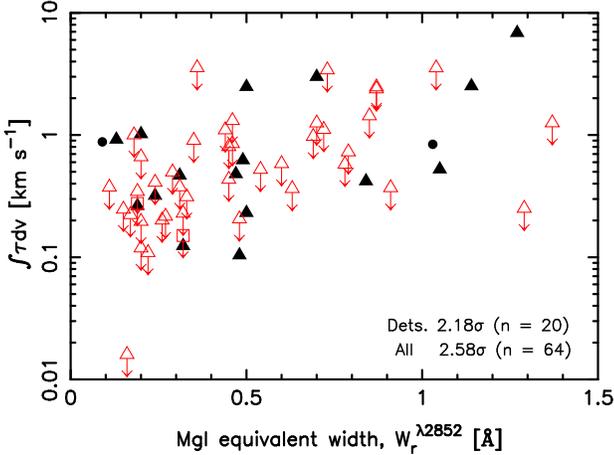}
\caption{As the left panel of Fig. \ref{strength_W_ratio}, but with the velocity
  integrated optical depth of the 21-cm line on the ordinate and the
  equivalent width of the Mg{\sc \,i} 2852 \AA\ line on the
  abscissa. Note that for ${\rm W}_{\rm r}^{\lambda2852}>0.5$ \AA,
  {\em all} of the 21-cm non-detections have $DA_{\rm abs}/DA_{\rm
    QSO}\geq0.85$.}
\label{2852_ratio}
\end{figure}
In Fig. \ref{2852_ratio} we extend this to the sample searched for
21-cm absorption. Ideally, we would show the normalised line strength
($\int\!\tau\,dv/N_{\rm HI}\propto f/T_{\rm spin}$, as in
Fig. \ref{strength_W_ratio}), however there are only 12 absorbers with
measurements of both (comprising of seven 21-cm detections and five
non-detections)\footnote{Although these few points do give a
  $2.21\sigma$ correlation between $\int\!\tau\,dv/N_{\rm HI}$ and
  ${\rm W}_{\rm r}^{\lambda2852}$.}. From this, we see a weak
correlation between $\int\!\tau\,dv$ and ${\rm W}_{\rm
  r}^{\lambda2852}$, although, as was noted in \citet{ctp+07}, the
neutral hydrogen column density is required in order to obtain the
full picture (and $f/T_{\rm spin}$). The absence of a measured $N_{\rm
  HI}$ could be responsible for the fairly weak correlation,
especially considering that the non-detections tend to have lower
column densities (Sect. \ref{rdng}).

Finally, for completeness, in Fig. \ref{2796_ratio} we show $\int\!\tau\,dv$ versus
the Mg{\sc \,ii} 2796 \AA\ equivalent width for the whole sample and see that,
\begin{figure}
\centering \includegraphics[angle=270,scale=0.75]{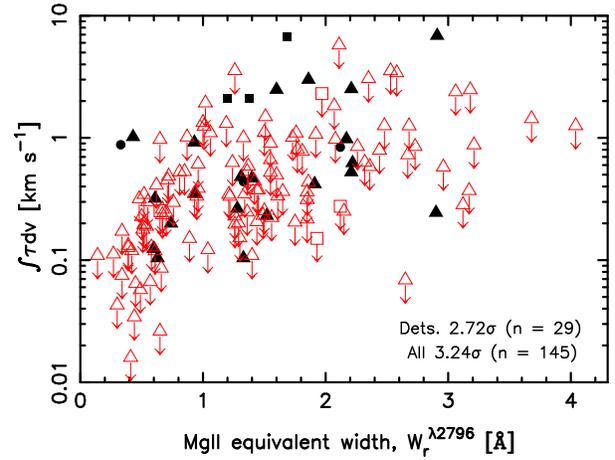}
\caption{As Fig. \ref{2852_ratio}, but with the equivalent width
  of the Mg{\sc \,ii} 2796 \AA\ line on the abscissa.}
\label{2796_ratio}
\end{figure}
although the correlation is negligible for the confirmed DLAs only \citep{ctp+07},
it becomes quite significant over a sample which is an order of magnitude larger.
Bear in mind, however, that the FWHMs for the 21-cm non-detections in Fig. \ref{2852_ratio} and
\ref{2796_ratio} have been estimated from the FWHM--${\rm W}_{\rm r}^{\lambda2796}-\,$[M/H]
correlations of the detections (Sect. \ref{pw}), which will tend to increase any
significance derived for these. This will be tempered somewhat by the 
optical depth limits for the non-detections, which are known. Therefore the significance
of each of these correlations should be considered to lie somewhere between the two 
quoted values. 

To summarise our findings regarding the detectability of 21-cm absorption in 
intervening Mg{\sc \,ii}  absorption systems:
\begin{itemize}

  \item On the basis of a 21-cm detection and non-detection being
    equally probable, at small angular diameter distance
    ratios ($DA_{\rm abs}/DA_{\rm QSO}\lapp0.8$) the observed
    distribution is expected for the whole sample, whereas at high ratios the
    observed distribution is very improbable. This suggests that the
    angular diameter distance ($DA_{\rm abs}$) plays a crucial
    r\^{ole} in the detection of 21-cm absorption, which in turn
    suggests a strong covering factor dependence.

  \item Large angular diameter distance ratios may also explain why
    the non-detections span a similar range of Mg{\sc \,ii} 2796
    \AA\ equivalent widths as the detections, which exhibit a 21-cm
    line strength--${\rm W}_{\rm
      r}^{\lambda2796}$ correlation. Note also, that, in general, the
    non-detections have considerably lower neutral hydrogen column
    densities, which could contribute to making these harder to detect over the whole
    Mg{\sc \,ii} sample\footnote{Although $N_{\rm HI}$ is normalised
      out in our definition of line strength in the case of the $\int\!\tau\,dv/N_{\rm HI}$---${\rm W}_{\rm
      r}^{\lambda2796}$ correlation.}.

    \item Using this correlation to estimate the column densities for
      the absorbers in which these are unmeasured (particularly at
      $1.1\lapp z_{\rm abs}\lapp1.6$), we find that the cosmological
      mass density of the neutral gas does not deviate significantly
      from $\Omega_{\rm neutral~gas}\approx1\times10^{-3}$ over the
      redshifts spanned by the optically selected absorbers ($0.1\lapp
      z_{\rm abs}\lapp3.5$). This consistency with the general
      optically selected population supports the 21-cm line
      strength--${\rm W}_{\rm r}^{\lambda2796}$ correlation.
\end{itemize}
%    \item 
\citet{ctp+07} suggested that the $\int\!\tau\,dv/N_{\rm HI}$---${\rm
  W}_{\rm r}^{\lambda2796}$ correlation was dominated by the 21-cm
profile width in confirmed DLAs, although this does not appear to be
the case for the Mg{\sc \,ii} sample in general
\citep{gsp+09a,kpec09}.  However, we find this significance to
increase when only the absorbers with Mg{\sc \,i} 2852 \AA\ equivalent
widths of ${\rm W}_{\rm r}^{\lambda2852}>0.5$ \AA\ are added. This
happens to be a sub-set of the Mg{\sc \,ii} absorbers with $1 < {\rm
  W}_{\rm r}^{\lambda2796}/{\rm W}_{\rm r}^{\lambda2600}<2$ {\sc and}
${\rm W}_{\rm r}^{\lambda2852}>0.1$ \AA, $\approx40\%$ of which are
known to be DLAs \citep{rtn05}. The FWHM--${\rm W}_{\rm
  r}^{\lambda2796}$ correlation does therefore appear to hold for
absorbers in which the neutral hydrogen column density exceeds $N_{\rm
  HI}\sim10^{20}$ \scm. Finally, we note a (somewhat scattered)
correlation between the total neutral hydrogen column density and the
equivalent width of the Mg{\sc \,i} 2852 \AA\ line, although there
appears to be no such relation for the singly ionised metal species
\citep{pdlc04}.

% \item Although not evident from a sample of 13 \citep{ctp+07}, for all of the Mg{\sc \,ii} absorbers searched, there is a correlation between the 21-cm line strength and the Mg{\sc \,ii} 2796 \AA\  equivalent width.

%\end{itemize}

\section*{Acknowledgments}

I wish to thank the referee for their helpful comments, which aided in
clarifying the manuscript.  This research has made use of the NASA/IPAC
Extragalactic Database (NED) which is operated by the Jet Propulsion
Laboratory, California Institute of Technology, under contract with
the National Aeronautics and Space Administration.  This research has
also made use of NASA's Astrophysics Data System Bibliographic Service
and {\sc asurv} Rev 1.2 \citep{lif92a}, which implements the methods
presented in \citet{ifn86}.

%\bibliographystyle{mn2e}
%\bibliography{aa,ref}

\begin{thebibliography}{}

\bibitem[\protect\citeauthoryear{Briggs \& Wolfe}{Briggs \& Wolfe}{1983}]{bw83}
Briggs F.~H.,  Wolfe A.~M.,  1983, ApJ, 268, 76

\bibitem[\protect\citeauthoryear{{Brown} \& {Mitchell}}{{Brown} \&
  {Mitchell}}{1983}]{bm83}
{Brown} R.~L.,  {Mitchell} K.~J.,  1983, ApJ, 264, 87

\bibitem[\protect\citeauthoryear{{Brown} \& {Roberts}}{{Brown} \&
  {Roberts}}{1973}]{br73}
{Brown} R.~L.,  {Roberts} M.~S.,  1973, ApJ, 184, L7

\bibitem[\protect\citeauthoryear{{Carilli}, {Lane}, {de Bruyn}, {Braun} \&
  {Miley}}{{Carilli} et~al.}{1996}]{cld+96}
{Carilli} C.~L.,  {Lane} W.,  {de Bruyn} A.~G.,  {Braun} R.,    {Miley} G.~K.,
  1996, AJ, 111, 1830

\bibitem[\protect\citeauthoryear{{Carilli}, {Rupen} \& {Yanny}}{{Carilli}
  et~al.}{1993}]{cry93}
{Carilli} C.~L.,  {Rupen} M.~P.,    {Yanny} B.,  1993, ApJ, 412, L59

\bibitem[\protect\citeauthoryear{{Chengalur}, {de Bruyn} \&
  {Narasimha}}{{Chengalur} et~al.}{1999}]{cdn99}
{Chengalur} J.~N.,  {de Bruyn} A.~G.,    {Narasimha} D.,  1999, A\&A, 343, L79

\bibitem[\protect\citeauthoryear{{Chengalur} \& {Kanekar}}{{Chengalur} \&
  {Kanekar}}{2000}]{ck00}
{Chengalur} J.~N.,  {Kanekar} N.,  2000, MNRAS, 318, 303

\bibitem[\protect\citeauthoryear{Curran, Darling, Bolatto, Whiting, Bignell \&
  Webb}{Curran et~al.}{2007a}]{cdbw07}
Curran S.~J.,  Darling J.~K.,  Bolatto A.~D.,  Whiting M.~T.,  Bignell C.,
  Webb J.~K.,  2007a, MNRAS, 382, L11

\bibitem[\protect\citeauthoryear{Curran, Kanekar \& Darling}{Curran
  et~al.}{2004}]{cdk04}
Curran S.~J.,  Kanekar N.,    Darling J.~K.,  2004, Science with the Square
  Kilometer Array, New Astronomy Reviews 48.
Elsevier, Amsterdam, pp 1095--1105

\bibitem[\protect\citeauthoryear{Curran, Murphy, Pihlstr\"{o}m, Webb \&
  Purcell}{Curran et~al.}{2005}]{cmp+03}
Curran S.~J.,  Murphy M.~T.,  Pihlstr\"{o}m Y.~M.,  Webb J.~K.,    Purcell
  C.~R.,  2005, MNRAS, 356, 1509

\bibitem[\protect\citeauthoryear{Curran, Tzanavaris, Darling, Whiting, Webb,
  Bignell, Athreya \& Murphy}{Curran et~al.}{2009}]{ctd+09}
Curran S.~J.,  Tzanavaris P.,  Darling J.~K.,  Whiting M.~T.,  Webb J.~K.,
  Bignell C.,  Athreya R.,    Murphy M.~T.,  2009, MNRAS, In press (arXiv:0910.3742)

\bibitem[\protect\citeauthoryear{Curran, Tzanavaris, Murphy, Webb \&
  Pihlstr\"{o}m}{Curran et~al.}{2007b}]{ctm+07}
Curran S.~J.,  Tzanavaris P.,  Murphy M.~T.,  Webb J.~K.,    Pihlstr\"{o}m
  Y.~M.,  2007b, MNRAS, 381, L6

\bibitem[\protect\citeauthoryear{Curran, Tzanavaris, Pihlstr\"{o}m \&
  Webb}{Curran et~al.}{2007c}]{ctp+07}
Curran S.~J.,  Tzanavaris P.,  Pihlstr\"{o}m Y.~M.,    Webb J.~K.,  2007c,
  MNRAS, 382, 1331

\bibitem[\protect\citeauthoryear{Curran \& Webb}{Curran \& Webb}{2006}]{cw06}
Curran S.~J.,  Webb J.~K.,  2006, MNRAS, 371, 356

\bibitem[\protect\citeauthoryear{Curran, Whiting, Wiklind, Webb, Murphy \&
  Purcell}{Curran et~al.}{2008}]{cww+08}
Curran S.~J.,  Whiting M.~T.,  Wiklind T.,  Webb J.~K.,  Murphy M.~T.,
  Purcell C.~R.,  2008, MNRAS, 391, 765

\bibitem[\protect\citeauthoryear{{Darling}, {Giovanelli}, {Haynes}, {Bower} \&
  {Bolatto}}{{Darling} et~al.}{2004}]{dgh+04}
{Darling} J.,  {Giovanelli} R.,  {Haynes} M.~P.,  {Bower} G.~C.,    {Bolatto}
  A.~D.,  2004, ApJ, 613, L101

\bibitem[\protect\citeauthoryear{{de Bruyn}, {O'Dea} \& {Baum}}{{de Bruyn}
  et~al.}{1996}]{dob96}
{de Bruyn} A.~G.,  {O'Dea} C.~P.,    {Baum} S.~A.,  1996, A\&A, 305, 450

\bibitem[\protect\citeauthoryear{{Ellison}}{{Ellison}}{2006}]{ell06}
{Ellison} S.~L.,  2006, MNRAS, 368, 335

\bibitem[\protect\citeauthoryear{{Gupta}, {Srianand}, {Petitjean}, {Noterdaeme}
  \& {Saikia}}{{Gupta} et~al.}{2009}]{gsp+09a}
{Gupta} N.,  {Srianand} R.,  {Petitjean} P.,  {Noterdaeme} P.,    {Saikia}
  D.~J.,  2009, MNRAS, 398, 201

\bibitem[\protect\citeauthoryear{{Isobe}, {Feigelson} \& {Nelson}}{{Isobe}
  et~al.}{1986}]{ifn86}
{Isobe} T.,  {Feigelson} E.,    {Nelson} P.,  1986, ApJ, 306, 490

\bibitem[\protect\citeauthoryear{Kanekar \& Briggs}{Kanekar \&
  Briggs}{2003}]{kb03}
Kanekar N.,  Briggs F.~H.,  2003, A\&A, 412, L29

\bibitem[\protect\citeauthoryear{Kanekar \& Chengalur}{Kanekar \&
  Chengalur}{2001}]{kc01a}
Kanekar N.,  Chengalur J.~N.,  2001, A\&A, 369, 42

\bibitem[\protect\citeauthoryear{Kanekar \& Chengalur}{Kanekar \&
  Chengalur}{2003}]{kc02}
Kanekar N.,  Chengalur J.~N.,  2003, A\&A, 399, 857

\bibitem[\protect\citeauthoryear{{Kanekar}, {Chengalur}, {Subrahmanyan} \&
  {Petitjean}}{{Kanekar} et~al.}{2001}]{kcsp01}
{Kanekar} N.,  {Chengalur} J.~N.,  {Subrahmanyan} R.,    {Petitjean} P.,  2001,
  A\&A, 367, 46

\bibitem[\protect\citeauthoryear{{Kanekar}, {Lane}, {Momjian}, {Briggs} \&
  {Chengalur}}{{Kanekar} et~al.}{2009a}]{klm+09}
{Kanekar} N.,  {Lane} W.~M.,  {Momjian} E.,  {Briggs} F.~H.,    {Chengalur}
  J.~N.,  2009a, MNRAS, 394, L61

\bibitem[\protect\citeauthoryear{{Kanekar}, {Prochaska}, {Ellison} \&
  {Chengalur}}{{Kanekar} et~al.}{2009b}]{kpec09}
{Kanekar} N.,  {Prochaska} J.~X.,  {Ellison} S.~L.,    {Chengalur} J.~N.,
  2009b, MNRAS, 396, 385

\bibitem[\protect\citeauthoryear{{Kanekar}, {Subrahmanyan}, {Ellison}, {Lane}
  \& {Chengalur}}{{Kanekar} et~al.}{2006}]{kse+06}
{Kanekar} N.,  {Subrahmanyan} R.,  {Ellison} S.~L.,  {Lane} W.,    {Chengalur}
  J.~N.,  2006, MNRAS, 370, L46

\bibitem[\protect\citeauthoryear{Lane, Smette, Briggs, Rao, Turnshek \&
  Meylan}{Lane et~al.}{1998}]{lsb+98}
Lane W.,  Smette A.,  Briggs F.~H.,  Rao S.~M.,  Turnshek D.~A.,    Meylan G.,
  1998, AJ, 116, 26

\bibitem[\protect\citeauthoryear{Lane}{Lane}{2000}]{lan00}
Lane W.~M.,  2000, PhD thesis, University of Groningen

\bibitem[\protect\citeauthoryear{Lane \& Briggs}{Lane \& Briggs}{2001}]{lb01}
Lane W.~M.,  Briggs F.~H.,  2001, ApJ, 561, L27

\bibitem[\protect\citeauthoryear{{Lane}, {Briggs} \& {Smette}}{{Lane}
  et~al.}{2000}]{lbs00}
{Lane} W.~M.,  {Briggs} F.~H.,    {Smette} A.,  2000, ApJ, 532, 146

\bibitem[\protect\citeauthoryear{{Lavalley}, {Isobe} \& {Feigelson}}{{Lavalley}
  et~al.}{1992}]{lif92a}
{Lavalley} M.~P.,  {Isobe} T.,    {Feigelson} E.~D.,  1992, in BAAS Vol.~24,
  {ASURV, Pennsylvania State University. Report for the period Jan 1990 - Feb
  1992.}.
pp 839--840

\bibitem[\protect\citeauthoryear{{Lovell}, {Reynolds}, {Jauncey}, {Backus},
  {McCulloch}, {Sinclair}, {Wilson}, {Tzioumis}, {King}, {Gough}, {Ellingsen},
  {Phillips}, {Preston} \& {Jones}}{{Lovell} et~al.}{1996}]{lrj+96}
{Lovell} J.~E.~J.,  {Reynolds} J.~E.,  {Jauncey} D.~L., et al, 1996, ApJ, 472, L5

\bibitem[\protect\citeauthoryear{{P\'{e}roux}, {Deharveng}, {Le Brun} \&
  {Cristiani}}{{P\'{e}roux} et~al.}{2004}]{pdlc04}
{P\'{e}roux} C.,  {Deharveng} J.-M.,  {Le Brun} V.,    {Cristiani} S.,  2004,
  MNRAS, 352, 1291

\bibitem[\protect\citeauthoryear{{Pihlstr{\" o}m}, {Vermeulen}, {Taylor} \&
  {Conway}}{{Pihlstr{\" o}m} et~al.}{1999}]{pvtc99}
{Pihlstr{\" o}m} Y.~M.,  {Vermeulen} R.~C.,  {Taylor} G.~B.,    {Conway} J.~E.,
   1999, ApJ, 525, L13

\bibitem[\protect\citeauthoryear{Prochaska \& Herbert-Fort}{Prochaska \&
  Herbert-Fort}{2004}]{ph04}
Prochaska J.~X.,  Herbert-Fort S.,  2004, PASP, 116, 622

\bibitem[\protect\citeauthoryear{Rao, Turnshek \& Nestor}{Rao
  et~al.}{2006}]{rtn05}
Rao S.,  Turnshek D.,    Nestor D.~B.,  2006, ApJ, 636, 610

\bibitem[\protect\citeauthoryear{{Rao} \& {Turnshek}}{{Rao} \&
  {Turnshek}}{2000}]{rt00}
{Rao} S.~M.,  {Turnshek} D.~A.,  2000, ApJS, 130, 1

\bibitem[\protect\citeauthoryear{{Roberts}, {Brown}, {Brundage}, {Rots},
  {Haynes} \& {Wolfe}}{{Roberts} et~al.}{1976}]{rbb+76}
{Roberts} M.~S.,  {Brown} R.~L.,  {Brundage} W.~D.,  {Rots} A.~H.,  {Haynes}
  M.~P.,    {Wolfe} A.~M.,  1976, AJ, 81, 293

\bibitem[\protect\citeauthoryear{Wolfe, Briggs \& Jauncey}{Wolfe
  et~al.}{1981}]{wbj81}
Wolfe A.~M.,  Briggs F.~H.,    Jauncey D.~L.,  1981, ApJ, 248, 460

\bibitem[\protect\citeauthoryear{{Wolfe}, {Briggs}, {Turnshek}, {Davis},
  {Smith} \& {Cohen}}{{Wolfe} et~al.}{1985}]{wbt+85}
{Wolfe} A.~M.,  {Briggs} F.~H.,  {Turnshek} D.~A.,  {Davis} M.~M.,  {Smith}
  H.~E.,    {Cohen} R.~D.,  1985, ApJ, 294, L67

\bibitem[\protect\citeauthoryear{Wolfe \& Davis}{Wolfe \& Davis}{1979}]{wd79}
Wolfe A.~M.,  Davis M.~M.,  1979, AJ, 84, 699

\bibitem[\protect\citeauthoryear{York, Kanekar, Ellison \& Pettini}{York
  et~al.}{2007}]{ykep07}
York B.~A.,  Kanekar N.,  Ellison S.~L.,    Pettini M.,  2007, MNRAS, 382, L53

\end{thebibliography}

\label{lastpage}

\end{document}